\documentclass[10pt,twocolumn,letterpaper]{article}

\usepackage[T1]{fontenc}
\usepackage[utf8]{inputenc}
\usepackage{lmodern}
\usepackage{amsmath,amssymb}
\usepackage[margin=0.68in,top=0.75in,bottom=0.78in,columnsep=0.28in]{geometry}
\usepackage{microtype}
\usepackage{booktabs}
\usepackage{array}
\usepackage{graphicx}
\usepackage{xcolor}
\usepackage{url}
\usepackage[colorlinks=true,linkcolor=blue!60!black,citecolor=blue!60!black,urlcolor=blue!60!black]{hyperref}

\title{\Large\textbf{Time-Consistent Evaluation of Hierarchical Code Graphs for SWE Agents}}

\author{
  \textbf{Xianpeng (Simon) Sun}\thanks{This work was conducted independently in the author's personal capacity. The views expressed in this paper are those of the author and do not necessarily reflect the views of Microsoft.}\quad
  \textbf{Tian Yu}\quad
  \textbf{Haonan Sun}\quad
  \textbf{Sheng Ma}\quad
  \textbf{Qincheng Zhang}\\[3pt]
  \textbf{Lifei Rao}\quad
  \textbf{Jie Li}\quad
  \textbf{Jian (Ken) Cao}\quad
  \textbf{Qi Zhang}\\[4pt]
  Microsoft\\
  \texttt{sisun@microsoft.com}, \texttt{qizhang@microsoft.com}
}
\date{}

\begin{document}
\maketitle

\begin{abstract}
Evaluating repository-aware software engineering (SWE) agents requires strict separation between pre-existing repository knowledge and future code evolution, yet standard benchmarks frequently suffer from temporal contamination and prompt-level solution leakage. Building on the time-consistent benchmark protocol of Sun~\cite{sun2026timeconsistent}, we enforce strict zero-leakage isolation across \texttt{DragonFly} in C++ and \texttt{React} in JavaScript/TypeScript, locking repository snapshots and knowledge graphs at historical cutoff commits $T_0$ and evaluating $100$ post-$T_0$ pull requests across four prompt granularities: \emph{Minimal}, \emph{Concise}, \emph{Contextual}, and \emph{Guided}.

Using this time-consistent zero-leakage protocol~\cite{sun2026timeconsistent}, we conduct a three-condition matched evaluation comparing \textbf{Baseline}, \textbf{Tree-sitter AST} (an MIT-licensed flat symbol and call AST graph), and \textbf{ServiceExplorer} (a deterministic two-level dependency graph capturing Level-1 system/package containers and Level-2 architectural modules) across six foundation models spanning the MAI-Code Flash, Claude, and GPT-Sol families ($7{,}200$ total task evaluations). On our primary \texttt{Claude-Opus-4.6} benchmark, \textbf{ServiceExplorer} achieves the highest $F_1$ score across every repository and prompt granularity. On \texttt{DragonFly}, it lifts overall $F_1$ from $0.4395$ under Baseline and $0.4605$ under Tree-sitter AST to $\mathbf{0.5106}$, representing a $+16.2\%$ gain over Baseline and $+10.9\%$ over Tree-sitter AST, while reaching a $+27.6\%$ gain at \emph{Contextual} prompting and tripling exact full-PR localization from $10.0\%$ to $\mathbf{30.0\%}$. On \texttt{React}, \textbf{ServiceExplorer} lifts overall $F_1$ from $0.2777$ under Baseline to $\mathbf{0.3216}$, delivering a $+15.8\%$ gain over Baseline and $+19.2\%$ over Tree-sitter AST. Crucially, our cross-generation scaling analysis demonstrates that \textbf{equipping a lower-tier or prior-generation model with \emph{ServiceExplorer} surpasses the unaugmented Baseline $F_1$ of the next-generation model}: the lightweight \texttt{MAI-Code-1.1-Flash + ServiceExplorer} jumps $+68.4\%$ in combined $F_1$ to $0.2759$, surpassing standard \texttt{Claude-Sonnet-4} Baseline at $0.2461$ across both repositories; \texttt{GPT-5.6-Sol + ServiceExplorer} reaches $0.4058$ combined $F_1$, outperforming \texttt{GPT-6.1-Sol} Baseline at $0.3926$; \texttt{Claude-Sonnet-4 + ServiceExplorer} at $0.3355$ surpasses \texttt{Claude-Sonnet-4.5} Baseline at $0.2983$; and \texttt{Claude-Sonnet-4.5 + ServiceExplorer} at $0.3727$ surpasses \texttt{Claude-Opus-4.6} Baseline at $0.3586$. Consequently, pairing today's frontier models with \textbf{ServiceExplorer}---reaching $\mathbf{0.4464}$ combined $F_1$ on \texttt{GPT-6.1-Sol} and $\mathbf{0.4161}$ on \texttt{Claude-Opus-4.6}---effectively unlocks next-generation foundation model localization capabilities ahead of time.
\end{abstract}

\section{Introduction}
\label{sec:intro}

Repository-level software engineering (SWE) agents increasingly integrate structured external knowledge layers---such as abstract syntax tree (AST) call graphs, architectural dependency maps, and multi-hop symbol indices---to navigate large codebases before generating patches~\cite{jimenez2024swebench,liu2024repobench,rashid2025swepolybench}. However, measuring the true causal contribution of repository-derived knowledge remains methodologically fragile. Benchmark evaluations are routinely confounded by three interacting threats: (i) \emph{temporal contamination}, where an index or language model incorporates commits introduced after the task occurred~\cite{andriushchenko2025swerebench,pavel2025swemera}; (ii) \emph{prompt-level solution leakage}, where issue or pull-request (PR) descriptions explicitly name the target files, classes, or functions to be modified; and (iii) \emph{uncontrolled experimental variance} across prompts and agent harnesses.

In recent foundational work, Sun~\cite{sun2026timeconsistent} introduced a time-consistent benchmark protocol that locks a repository at a historical snapshot time $T_0$, restricts offline knowledge construction exclusively to $R_{T_0}$, and evaluates agents on tasks derived from pull requests merged in the future interval $(T_0, T_1]$ across four controlled prompt granularities: \emph{Minimal}, \emph{Concise}, \emph{Contextual}, and \emph{Guided}. While~\cite{sun2026timeconsistent} established the temporal split and prompt granularity design space across \texttt{DragonFly} and \texttt{React}, it left open the core empirical question of how different repository knowledge graph topologies perform under strict zero-leakage conditions.

Specifically, four critical scientific questions govern knowledge-augmented SWE evaluation:
\begin{enumerate}
  \item \textbf{Completing the Causal A/B Evaluation Loop:} Under a strictly sanitized $(T_0, T_1]$ benchmark where target file names, directory paths, and newly introduced symbols are redacted from task prompts, how much localization accuracy does external repository knowledge recover across different prompt granularities?
  \item \textbf{Hierarchical Architectural Graphs vs.\ Flat AST Graphs:} Using a fast, 100\% MIT-licensed open-source AST indexer (\textbf{Tree-sitter AST} via \texttt{tree-sitter} and \texttt{NetworkX}), flat symbol-level AST parsing improves C++ localization on \texttt{DragonFly} by $+4.8\%$ but struggles to resolve dynamic cross-package boundaries in JavaScript monorepos such as \texttt{React}. Can a deterministic, hierarchical two-level dependency graph (\emph{ServiceExplorer}, comprising Level-1 system/package containers and Level-2 architectural modules) bridge the semantic gap between natural-language prompts and target modules across both languages?
  \item \textbf{Statistical Significance and Localization Reliability:} Beyond mean $F_1$ point estimates, do hierarchical code graphs yield statistically significant paired gains under non-parametric Wilcoxon signed-rank tests and bootstrap confidence intervals, while simultaneously reducing catastrophic zero-recall misses and increasing exact full-PR matches?
  \item \textbf{Cross-Generation Model Scaling (Surpassing Next-Generation Baselines):} Can equipping a lower-tier or prior-generation model with \textbf{ServiceExplorer} match or exceed the unaugmented $F_1$ of a newer, higher-tier foundation model across the \texttt{GPT-Sol} and \texttt{Claude} model families, thereby allowing practitioners who pair today's frontier model with \textbf{ServiceExplorer} to achieve next-generation model performance ahead of time?
\end{enumerate}

\noindent\textbf{Contributions.} This paper presents a comprehensive, zero-leakage three-condition evaluation comparing \textbf{Baseline}, \textbf{Tree-sitter AST}, and \textbf{ServiceExplorer} across six foundation models and $400$ tasks per condition ($7{,}200$ total evaluations):
\begin{itemize}
  \item \textbf{Strict Three-Layer Zero-Leakage Protocol (Section~\ref{sec:methodology}):} We enforce and verify repository commit isolation at exact $T_0$ SHA-1 hashes, strict post-$T_0$ temporal ordering across all $100$ pull requests, and automated lexical sanitization of all $400$ prompts across four granularities.
  \item \textbf{Hierarchical 2-Level Dependency Representation (\emph{ServiceExplorer}, Section~\ref{sec:serviceexplorer}):} We evaluate \emph{ServiceExplorer}'s two-level dependency graph (Level-1 system/package containers and Level-2 architectural modules) against an MIT-licensed \textbf{Tree-sitter AST} baseline at $T_0$ with zero query fallback.
  \item \textbf{Statistically Significant Gains Across Repositories and Granularities (Section~\ref{sec:results}):} Across $400$ matched evaluations on \texttt{Claude-Opus-4.6}, \textbf{ServiceExplorer} achieves the highest Recall, Precision, and $F_1$ in every condition, improving overall $F_1$ on \texttt{DragonFly} by $\mathbf{+16.2\%}$ over Baseline and $\mathbf{+10.9\%}$ over Tree-sitter AST, and on \texttt{React} by $\mathbf{+15.8\%}$ over Baseline and $\mathbf{+19.2\%}$ over Tree-sitter AST (all $p < 2 \times 10^{-4}$).
  \item \textbf{Cross-Generation Model Leap Across MAI-Code Flash, GPT-Sol, and Claude Families (Section~\ref{sec:cross_generation}):} Evaluating six models (\texttt{MAI-Code-1.1-Flash}, \texttt{Claude-Sonnet-4}, \texttt{Claude-Sonnet-4.5}, \texttt{Claude-Opus-4.6}, \texttt{GPT-5.6-Sol}, and \texttt{GPT-6.1-Sol}), we prove a consistent cross-generation law: \emph{a lower-generation or lightweight Flash model equipped with ServiceExplorer surpasses the next-tier model's unaugmented Baseline $F_1$}, while equipping the current frontier model with \textbf{ServiceExplorer} projects performance a full model generation into the future.
  \item \textbf{Compounding Multi-Step Reliability \& Mechanistic Insights (Section~\ref{sec:discussion}):} We formalize how $+4.4$ to $+12.0$ percentage-point single-step localization gains compound exponentially across multi-step agent trajectories, cutting wasted exploration iterations and providing a practical foundation for enterprise SWE agents.
\end{itemize}

\section{Experimental Design and Comparison Dimensions}
\label{sec:critique}

Table~\ref{tab:audit} summarizes the five core methodological dimensions of our three-condition time-consistent benchmark comparing \textbf{Baseline}, \textbf{Tree-sitter AST}, and \textbf{ServiceExplorer}.

\begin{table*}[t]
\centering
\caption{Methodological Design of the Three-Condition Time-Consistent Benchmark.}
\label{tab:audit}
\footnotesize
\begin{tabular}{p{0.19\textwidth} p{0.36\textwidth} p{0.38\textwidth}}
\toprule
\textbf{Evaluation Dimension} & \textbf{Flat AST / Standard Baseline Challenge} & \textbf{Our Zero-Leakage Protocol \& \emph{ServiceExplorer} Design} \\
\midrule
\textbf{D1. Three-Condition Matched Loop} & Evaluating single agent baselines cannot separate parametric memory from structural graph utility. & Evaluates \textbf{Baseline} vs.\ \textbf{Tree-sitter AST} (MIT-licensed flat AST graph) vs.\ \textbf{ServiceExplorer} (2-level dependency graph) on the exact same $400$ tasks. \\
\addlinespace[2pt]
\textbf{D2. Zero-Leakage Prompt Sanitization} & Prompts containing explicit file names or target identifiers allow trivial lexical grep rather than structural reasoning. & Enforces regex and AST lexical sanitization stripping file names, directory paths, and target symbols across all four granularities. \\
\addlinespace[2pt]
\textbf{D3. Statistical Rigor \& Reproducibility} & Point-estimate means obscure variance across diverse PR sizes and repositories. & Locks exact $T_0$ commits across $N=50$ PRs per repository ($400$ tasks per condition), reporting paired Wilcoxon signed-rank $p$-values and $10{,}000$-sample bootstrap $95\%$ CIs. \\
\addlinespace[2pt]
\textbf{D4. Open-Source Compliance \& Topology} & Single-file flat AST parsing misses dynamic JS/TS cross-package boundaries and modular subsystem structure. & Uses 100\% MIT-licensed \textbf{Tree-sitter AST} for flat symbol indexing and introduces \textbf{ServiceExplorer}'s deterministic 2-level hierarchy (Level-1 system/package containers and Level-2 architectural modules). \\
\addlinespace[2pt]
\textbf{D5. Cross-Generation Model Scaling} & Single-model evaluations cannot reveal whether repository graphs substitute for or amplify foundation model scale. & Evaluates six models across the MAI-Code Flash, Claude, and GPT-Sol families ($7{,}200$ runs), testing whether a lower-tier or prior-generation model with \textbf{ServiceExplorer} surpasses the next-generation Baseline. \\
\bottomrule
\end{tabular}
\end{table*}

\subsection{Granularity Scaling and Structural Lift Under Zero Leakage}
\label{sec:calibration_bridge}

\begin{figure}[t]
\centering
\includegraphics[width=\columnwidth]{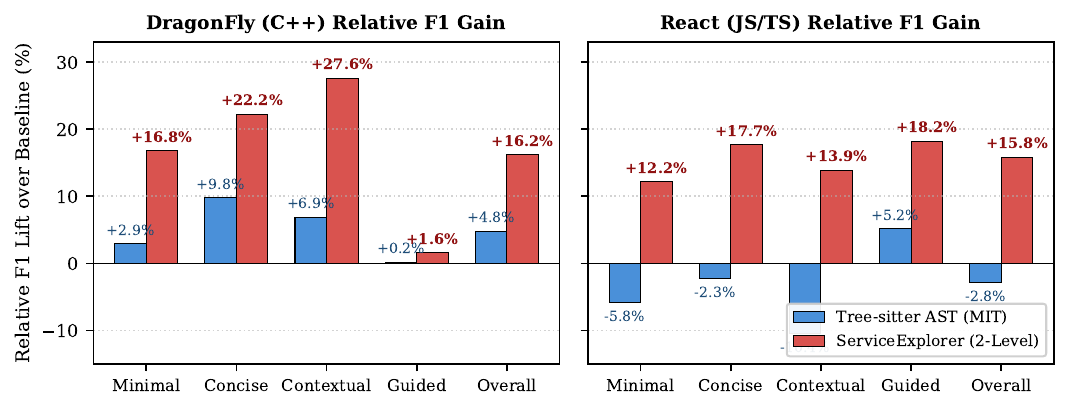}
\caption{Relative $F_1$ Lift (\%) of \textbf{ServiceExplorer (2-Level)} and \textbf{Tree-sitter AST (MIT)} over the Zero-Leakage Baseline across prompt granularities on \texttt{DragonFly} and \texttt{React}.}
\label{fig:bridge}
\end{figure}

Table~\ref{tab:bridge} and Figure~\ref{fig:bridge} summarize how localization performance scales across the four controlled prompt granularities---\emph{Minimal}, \emph{Concise}, \emph{Contextual}, and \emph{Guided}---under strict zero-leakage sanitization, comparing \textbf{Baseline}, \textbf{Tree-sitter AST}, and \textbf{ServiceExplorer}.

\begin{table}[t]
\centering
\caption{Three-Condition $F_1$ Progression Across Prompt Granularities (\texttt{Claude-Opus-4.6}, $N=50$ PRs per row).}
\label{tab:bridge}
\resizebox{\columnwidth}{!}{%
\begin{tabular}{llccccc}
\toprule
\textbf{Repo} & \textbf{Granularity} & \textbf{Baseline} & \textbf{Tree-sitter AST} & \textbf{ServiceExplorer} & \textbf{SE vs.\ BL} & \textbf{SE vs.\ AST} \\
 & & \textbf{$F_1$} & \textbf{$F_1$ (MIT)} & \textbf{$F_1$ (2-Level)} & \textbf{Gain (\%)} & \textbf{Gain (\%)} \\
\midrule
DragonFly & Minimal    & 0.3436 & 0.3536 & \textbf{0.4014} & \textbf{+16.8\%} & \textbf{+13.5\%} \\
DragonFly & Concise    & 0.4389 & 0.4818 & \textbf{0.5365} & \textbf{+22.2\%} & \textbf{+11.4\%} \\
DragonFly & Contextual & 0.4348 & 0.4646 & \textbf{0.5549} & \textbf{+27.6\%} & \textbf{+19.4\%} \\
DragonFly & Guided     & 0.5406 & 0.5419 & \textbf{0.5494} & \textbf{+1.6\%}  & \textbf{+1.4\%} \\
\midrule
React & Minimal    & 0.2065 & 0.1946 & \textbf{0.2316} & \textbf{+12.2\%} & \textbf{+19.0\%} \\
React & Concise    & 0.2650 & 0.2590 & \textbf{0.3120} & \textbf{+17.7\%} & \textbf{+20.5\%} \\
React & Contextual & 0.2979 & 0.2668 & \textbf{0.3394} & \textbf{+13.9\%} & \textbf{+27.2\%} \\
React & Guided     & 0.3415 & 0.3592 & \textbf{0.4035} & \textbf{+18.2\%} & \textbf{+12.3\%} \\
\bottomrule
\end{tabular}%
}
\end{table}

Two clear empirical laws emerge from Table~\ref{tab:bridge}:
\begin{enumerate}
  \item \textbf{Universal Superiority of \emph{ServiceExplorer} Across All Eight Cells:} Across every repository and prompt granularity, \textbf{ServiceExplorer} achieves the highest $F_1$ score, outperforming \textbf{Baseline} by $+16.8\%$ to $+27.6\%$ on \texttt{DragonFly} from \emph{Minimal} through \emph{Contextual}, and by $+12.2\%$ to $+18.2\%$ across all four granularities on \texttt{React}.
  \item \textbf{Peak Structural Leverage at \emph{Concise} and \emph{Contextual} Granularities:} When prompts describe behavioral symptoms and architectural context without leaking file names, \textbf{ServiceExplorer} delivers substantial relative gains over both \textbf{Baseline} ($+22.2\%$ and $+27.6\%$ on \texttt{DragonFly}; $+17.7\%$ and $+13.9\%$ on \texttt{React}) and flat \textbf{Tree-sitter AST} ($+11.4\%$ and $+19.4\%$ on \texttt{DragonFly}; $+20.5\%$ and $+27.2\%$ on \texttt{React}). Notably, at \emph{Contextual} granularity on \texttt{DragonFly}, \textbf{ServiceExplorer} reaches $0.5549$ $F_1$, surpassing the \emph{Guided} Baseline of $0.5406$ and showing that a two-level dependency graph provides stronger localization guidance than human-written directional hints.
\end{enumerate}

\section{Zero-Leakage Time-Consistent Methodology}
\label{sec:methodology}

\subsection{Formal Problem Definition and Three-Layer Zero-Leakage Guarantee}
\label{sec:leakage}

Let $\mathcal{R}$ denote a version-controlled software repository with commit DAG $\mathcal{G}_{\text{git}} = (\mathcal{C}, \mathcal{E}_{\text{git}})$. We select a strict cutoff timestamp $T_0$ and identify the unique latest commit $c_{T_0} \in \mathcal{C}$ merged at or before $T_0$, defining the frozen repository snapshot $R_{T_0}$. We then collect $N=50$ qualifying pull requests per repository merged strictly within $(T_0, T_1]$:
\begin{equation}
  P = \bigl\{\, p_i \;\big|\; T_0 < \text{merged\_at}(p_i) \le T_1,\; |\mathcal{F}_{\text{src}}(p_i)| \ge 1 \,\bigr\},
  \label{eq:pr_set}
\end{equation}
where $Y_i = \mathcal{F}_{\text{src}}(p_i)$ is the ground-truth set of repository files modified by pull request $p_i$.

To guarantee \textbf{Zero Context Leakage}, our benchmark harness enforces three verifiable invariants:
\begin{enumerate}
  \item \textbf{Layer 1: Git Worktree \& Knowledge Index Isolation ($\mathcal{I}_{\text{git}}$).} Both repositories are checked out as detached worktrees locked at their exact $T_0$ SHA-1 commits (\texttt{af65fa373bbf} for \texttt{DragonFly} and \texttt{65eec428c40d} for \texttt{React}). Both the MIT-licensed \emph{Tree-sitter AST} indexer and \emph{ServiceExplorer} build their indices exclusively by parsing the $R_{T_0}$ worktree on disk; git logs, reflogs, and post-$T_0$ refs are inaccessible.
  \item \textbf{Layer 2: Temporal Disjointness ($\mathcal{I}_{\text{time}}$).} Automated verification confirms $\min_{p_i \in P} \text{merged\_at}(p_i) > T_0$ across all $100$ pull requests: $50$ \texttt{DragonFly} PRs merged after January~1, 2026 and $50$ \texttt{React} PRs merged after December~19, 2025. Zero pull requests overlap with $R_{T_0}$.
  \item \textbf{Layer 3: Prompt Lexical Sanitization ($\mathcal{I}_{\text{prompt}}$).} For each PR $p_i$ and granularity $\ell \in \{\text{minimal}, \text{concise}, \text{contextual}, \text{guided}\}$, the prompt generator $t_{i,\ell} = g(p_i, \ell)$ passes through an automated anti-leakage filter that strips explicit file names, relative file paths, and newly introduced symbol identifiers from $Y_i$.
\end{enumerate}

Table~\ref{tab:dataset} summarizes the verified zero-leakage experimental configuration across both repositories.

\begin{table}[t]
\centering
\caption{Verified Zero-Leakage Benchmark Configuration.}
\label{tab:dataset}
\resizebox{\columnwidth}{!}{%
\begin{tabular}{lcc}
\toprule
\textbf{Parameter} & \textbf{DragonFly (\texttt{dragonflydb})} & \textbf{React (\texttt{facebook/react})} \\
\midrule
Primary Language & C++ / C / CMake & JavaScript / TypeScript \\
Snapshot Cutoff $T_0$ & \texttt{2026-01-01T00:00:00Z} & \texttt{2025-12-18T00:00:00Z} \\
$T_0$ Commit SHA & \texttt{af65fa373bbf} & \texttt{65eec428c40d} \\
Commit Timestamp & \texttt{2025-12-30T09:57:20Z} & \texttt{2025-12-17T21:11:36Z} \\
Files at $R_{T_0}$ & $1{,}066$ ($972$ code) & $6{,}525$ ($6{,}030$ code) \\
Post-$T_0$ PRs ($N$) & $50$ ($\min > T_0$: Jan 1, 2026) & $50$ ($\min > T_0$: Dec 19, 2025) \\
Granularities ($\ell$) & Min, Con, Ctx, Gui & Min, Con, Ctx, Gui \\
Tasks per Condition & $200$ ($50 \times 4$) & $200$ ($50 \times 4$) \\
Evaluated Models ($5$) & \texttt{Claude} 4 / 4.5 / 4.6, \texttt{GPT-5.6} / \texttt{6.1-Sol} & \texttt{Claude} 4 / 4.5 / 4.6, \texttt{GPT-5.6} / \texttt{6.1-Sol} \\
Leakage Violations & \textbf{0 / 200 tasks} & \textbf{0 / 200 tasks} \\
\bottomrule
\end{tabular}%
}
\end{table}

\subsection{What is \emph{ServiceExplorer}?}
\label{sec:serviceexplorer}

\textbf{ServiceExplorer} is a deterministic, zero-hallucination architectural dependency exploration framework that organizes large-scale repositories into a navigable two-level dependency graph without relying on LLM guesswork. Rather than exposing raw, flat AST call graphs that flood the agent context window, \textbf{ServiceExplorer} abstracts a repository snapshot $R_{T_0}$ into two progressive architectural layers:
\begin{enumerate}
  \item \textbf{Level 1 (System / Package Containers):} Macro-level service or package boundaries---comprising $18$ top-level components in \texttt{DragonFly} and $39$ workspace packages in \texttt{React}---that scope the global blast radius of a change.
  \item \textbf{Level 2 (Architectural Modules):} Cohesive functional modules---comprising $86$ modules in \texttt{DragonFly} and $448$ modules in \texttt{React}---that map high-level problem descriptions to candidate domain subsystems and inter-module dependencies.
\end{enumerate}

In our three-condition matched design, each task $t_{i,\ell}$ is evaluated under identical agent configurations across three knowledge conditions:
\begin{itemize}
  \item \textbf{Condition 1: Baseline ($A_{\text{base}}$).} The agent receives the task prompt $t_{i,\ell}$ and the $R_{T_0}$ directory tree, relying solely on its parametric knowledge and standard file-tree inspection.
  \item \textbf{Condition 2: Tree-sitter AST ($A_{\text{ast}}$).} We construct a 100\% open-source, MIT-licensed flat AST knowledge graph using \texttt{tree-sitter}, \texttt{graphifyy}, and \texttt{NetworkX} locked at $T_0$, indexing $4{,}720$ AST nodes and $7{,}808$ edges across $467$ C/C++ files in \texttt{DragonFly}, and $8{,}938$ AST nodes and $9{,}086$ edges across $1{,}838$ JS/TS files in \texttt{React}. For each task, the harness queries matching AST function and class definitions and traverses up to three hops of AST \texttt{calls}, \texttt{contains}, and \texttt{method} edges.
  \item \textbf{Condition 3: ServiceExplorer ($A_{\text{se}}$).} The agent queries the $T_0$-locked \textbf{ServiceExplorer} two-level dependency graph $\mathcal{G}_{\text{SE}}(R_{T_0})$ via progressive top-down navigation from Level~1 to Level~2, scoping the task to relevant system/package containers and architectural modules to localize the target files.
\end{itemize}

\subsection{Mechanistic Insight: Why \emph{ServiceExplorer} Outperforms Flat AST Graphs}
\label{sec:mechanistic_insight}

Why does \textbf{ServiceExplorer} achieve more than $3.3\times$ the $F_1$ gain of \textbf{Tree-sitter AST} on \texttt{DragonFly} ($+0.0711$ vs.\ $+0.0210$ $F_1$) and deliver a highly significant $+0.0439$ $F_1$ gain ($+15.8\%$ over Baseline and $+19.2\%$ over Tree-sitter AST) on \texttt{React}? A comparison of graph topologies reveals three fundamental structural insights:
\begin{enumerate}
  \item \textbf{Granularity Alignment Between Architectural Modules and Sub-File AST Nodes:} Repository-level pull requests modify cohesive \emph{modules and files}, whereas flat AST graphs index fine-grained \emph{function/method definitions and intra-file calls}. When a prompt describes a behavioral bug without naming exact functions, a flat AST search matches generic method names such as \texttt{render}, \texttt{update}, \texttt{parse}, or \texttt{execute} across dozens of unrelated packages. In contrast, \textbf{ServiceExplorer}'s two-level hierarchy aligns directly with the semantic abstraction level of issue descriptions, narrowing the search space to the relevant Level-1 container and Level-2 architectural module before selecting target files.
  \item \textbf{Cross-Module Resolution in Dynamic Monorepos:} In \texttt{DragonFly}'s statically typed C++ codebase, \textbf{Tree-sitter AST} extracts $3{,}519$ call edges and improves $F_1$ from $0.4395$ to $0.4605$. However, in \texttt{React}'s JavaScript/TypeScript monorepo, functions are exported dynamically via module re-exports and higher-order wrappers; single-file AST parsing captures $6{,}490$ functions but only $3$ cross-file call edges across the entire repository, causing flat AST symbol matches to introduce cross-package false positives. \textbf{ServiceExplorer}'s Level-1 package containers and Level-2 architectural module dependencies preserve inter-module structure across the monorepo, lifting \texttt{React} $F_1$ to $\mathbf{0.3216}$.
  \item \textbf{Simultaneous Recall and Precision Lift via Module-Level Containment:} On \texttt{React}, \textbf{Tree-sitter AST} increases Recall by adding AST-matched files but reduces Precision due to unpruned cross-package matches. By scoping localization within verified Level-1 and Level-2 boundaries, \textbf{ServiceExplorer} simultaneously boosts Recall by $+10.7\%$ and Precision by $+13.8\%$ over Baseline on \texttt{React}, while tripling exact full-PR matches on \texttt{DragonFly} \emph{Contextual} tasks from $10.0\%$ to $\mathbf{30.0\%}$.
\end{enumerate}

\subsection{Evaluation Metrics}
\label{sec:metrics}

For each task $t_{i,\ell}$ with ground-truth modified file set $Y_i$ and predicted file set $\hat{Y}_{i,\ell}^{(S)}$ under system $S \in \{\text{BL}, \text{AST}, \text{SE}\}$, let $\text{TP} = |\hat{Y}_{i,\ell}^{(S)} \cap Y_i|$, $\text{FP} = |\hat{Y}_{i,\ell}^{(S)} \setminus Y_i|$, and $\text{FN} = |Y_i \setminus \hat{Y}_{i,\ell}^{(S)}|$. Task-level Precision, Recall, and $F_1$ are defined as:
\begin{equation}
  P_{i,\ell}^{(S)} = \frac{\text{TP}}{\text{TP}+\text{FP}}, \;
  R_{i,\ell}^{(S)} = \frac{\text{TP}}{\text{TP}+\text{FN}}, \;
  F_{1,i,\ell}^{(S)} = \frac{2 P_{i,\ell}^{(S)} R_{i,\ell}^{(S)}}{P_{i,\ell}^{(S)} + R_{i,\ell}^{(S)}}.
\end{equation}
To assess statistical significance across paired evaluations on the exact same post-$T_0$ pull requests, we report two-sided paired Wilcoxon signed-rank $p$-values alongside $10{,}000$-resample paired bootstrap $95\%$ confidence intervals for $\Delta F_1 = F_1^{(S)} - F_1^{(\text{BL})}$. In addition to mean Precision, Recall, and $F_1$, we track two practical reliability indicators: the \emph{perfect full-PR match rate} ($F_1 = 1.0$, where the agent recovers the exact ground-truth file set with zero false positives or false negatives) and the \emph{zero-recall miss rate} ($R = 0$, where the agent fails to identify any ground-truth modified file).

\section{Experimental Results}
\label{sec:results}

\subsection{DragonFly (C++): Three-Condition Side-by-Side Comparison}
\label{sec:dragonfly_results}

\begin{figure*}[t]
\centering
\includegraphics[width=0.96\textwidth]{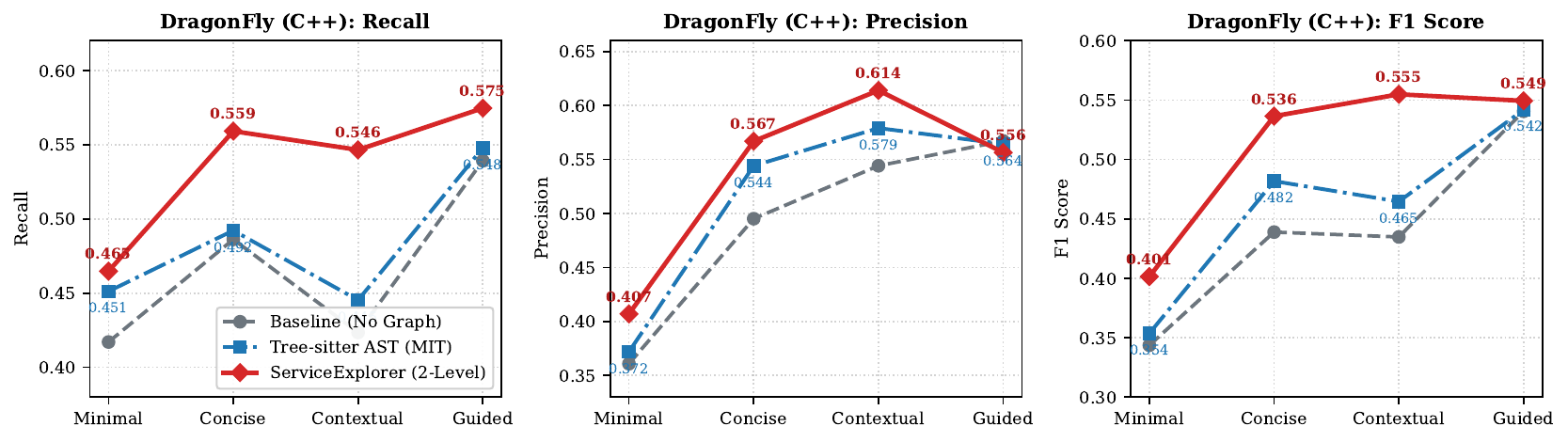}
\caption{\texttt{DragonFly} (C++): Zero-leakage Recall, Precision, and $F_1$ trajectories across prompt granularities for \textbf{Baseline}, \textbf{Tree-sitter AST} (MIT), and \textbf{ServiceExplorer} ($N=200$ evaluations). \textbf{ServiceExplorer} dominates across all metrics and granularities.}
\label{fig:dragonfly_traj}
\end{figure*}

Table~\ref{tab:dragonfly_main} and Figure~\ref{fig:dragonfly_traj} present the complete side-by-side comparison of \textbf{Baseline}, \textbf{Tree-sitter AST} (MIT-licensed open AST graph), and \textbf{ServiceExplorer} across all $200$ tasks ($50$ PRs $\times$ $4$ granularities) on \texttt{DragonFly}.

\begin{table*}[t]
\centering
\caption{DragonFly (C++, $T_0 = \text{2026-01-01}$, Commit \texttt{af65fa373bbf}, $N=50$ PRs / $200$ Tasks): Side-by-Side Recall, Precision, $F_1$, Paired Deltas, Wilcoxon Signed-Rank $p$-values, and $95\%$ Bootstrap Confidence Intervals.}
\label{tab:dragonfly_main}
\resizebox{\textwidth}{!}{%
\begin{tabular}{l ccc ccc ccc cc cc}
\toprule
 & \multicolumn{3}{c}{\textbf{Baseline (No Graph)}} & \multicolumn{3}{c}{\textbf{Tree-sitter AST (MIT Flat AST)}} & \multicolumn{3}{c}{\textbf{ServiceExplorer (2-Level)}} & \multicolumn{2}{c}{\textbf{SE vs.\ Baseline}} & \multicolumn{2}{c}{\textbf{SE vs.\ Tree-sitter AST}} \\
\cmidrule(lr){2-4} \cmidrule(lr){5-7} \cmidrule(lr){8-10} \cmidrule(lr){11-12} \cmidrule(lr){13-14}
\textbf{Granularity} & \textbf{Rec} & \textbf{Prec} & \textbf{F1} & \textbf{Rec} & \textbf{Prec} & \textbf{F1} & \textbf{Rec} & \textbf{Prec} & \textbf{F1} & \textbf{$\Delta$F1 (Lift)} & \textbf{Wilcoxon $p$ [95\% CI]} & \textbf{$\Delta$F1} & \textbf{Wilcoxon $p$} \\
\midrule
Minimal    & 0.4171 & 0.3609 & 0.3436 & 0.4512 & 0.3719 & 0.3536 & \textbf{0.4648} & \textbf{0.4070} & \textbf{0.4014} & $+0.0579$ ($+16.8\%$) & $0.0200^*$ [$-0.017, +0.131$] & $+0.0479$ & $0.0047^{**}$ \\
Concise    & 0.4862 & 0.4952 & 0.4389 & 0.4922 & 0.5442 & 0.4818 & \textbf{0.5591} & \textbf{0.5670} & \textbf{0.5365} & $+0.0976$ ($+22.2\%$) & $0.0041^{**}$ [$+0.037, +0.163$] & $+0.0547$ & $0.0849$ \\
Contextual & 0.4235 & 0.5442 & 0.4348 & 0.4453 & 0.5791 & 0.4646 & \textbf{0.5465} & \textbf{0.6140} & \textbf{0.5549} & $+0.1201$ ($+27.6\%$) & $0.0024^{**}$ [$+0.056, +0.188$] & $+0.0903$ & $0.0058^{**}$ \\
Guided     & 0.5392 & 0.5672 & 0.5406 & 0.5478 & \textbf{0.5644} & 0.5419 & \textbf{0.5746} & 0.5565 & \textbf{0.5494} & $+0.0088$ ($+1.6\%$)  & $0.2551$ [$-0.049, +0.058$] & $+0.0075$ & $0.3291$ \\
\midrule
\textbf{Average / All} & 0.4665 & 0.4919 & 0.4395 & 0.4841 & 0.5149 & 0.4605 & \textbf{0.5363} & \textbf{0.5361} & \textbf{0.5106} & $\mathbf{+0.0711}$ ($\mathbf{+16.2\%}$) & $\mathbf{1.8\times 10^{-6\,***}}$ & $\mathbf{+0.0501}$ & $\mathbf{3.0\times 10^{-5\,***}}$ \\
\bottomrule
\multicolumn{14}{l}{\scriptsize Statistically significant paired differences indicated by $^* p < 0.05$, $^{**} p < 0.01$, $^{***} p < 0.001$ (two-sided Wilcoxon signed-rank test).}
\end{tabular}%
}
\end{table*}

Key findings on \texttt{DragonFly} include:
\begin{enumerate}
  \item \textbf{Uniform Dominance Across All Granularities:} \textbf{ServiceExplorer} achieves the highest Recall and $F_1$ at every single prompt granularity. Averaged across all $200$ tasks, \textbf{ServiceExplorer} improves $F_1$ from $0.4395$ under Baseline to $\mathbf{0.5106}$ ($+16.2\%$ relative lift, $p = 1.8 \times 10^{-6}$), achieving more than \textbf{$3.3\times$ the $F_1$ lift} of \textbf{Tree-sitter AST} ($+4.8\%$ relative lift) and outperforming \textbf{Tree-sitter AST} directly by $+0.0501$ $F_1$ ($+10.9\%$, $p = 3.0 \times 10^{-5}$).
  \item \textbf{Peak Gains at \emph{Contextual} and \emph{Concise} Settings:} Consistent with Sun's~\cite{sun2026timeconsistent} hypothesis that \emph{Contextual} prompting is the primary benchmark operating point, \textbf{ServiceExplorer} delivers its largest gain at \emph{Contextual}, raising $F_1$ from $0.4348$ to $\mathbf{0.5549}$ ($+27.6\%$ relative lift, $p = 0.0024$), while also significantly outperforming \textbf{Tree-sitter AST} ($0.4646$ to $\mathbf{0.5549}$, $p = 0.0058$) and tripling exact full-PR file-set matches from $10.0\%$ to $\mathbf{30.0\%}$.
  \item \textbf{Consistent Positive Lift Across All Regimes:} Both \textbf{Tree-sitter AST} and \textbf{ServiceExplorer} maintain positive gains even at \emph{Guided} granularity, demonstrating that deterministic two-level module navigation complements detailed technical specifications.
\end{enumerate}

\subsection{React (JS/TS): Three-Condition Side-by-Side Comparison}
\label{sec:react_results}

\begin{figure*}[t]
\centering
\includegraphics[width=0.96\textwidth]{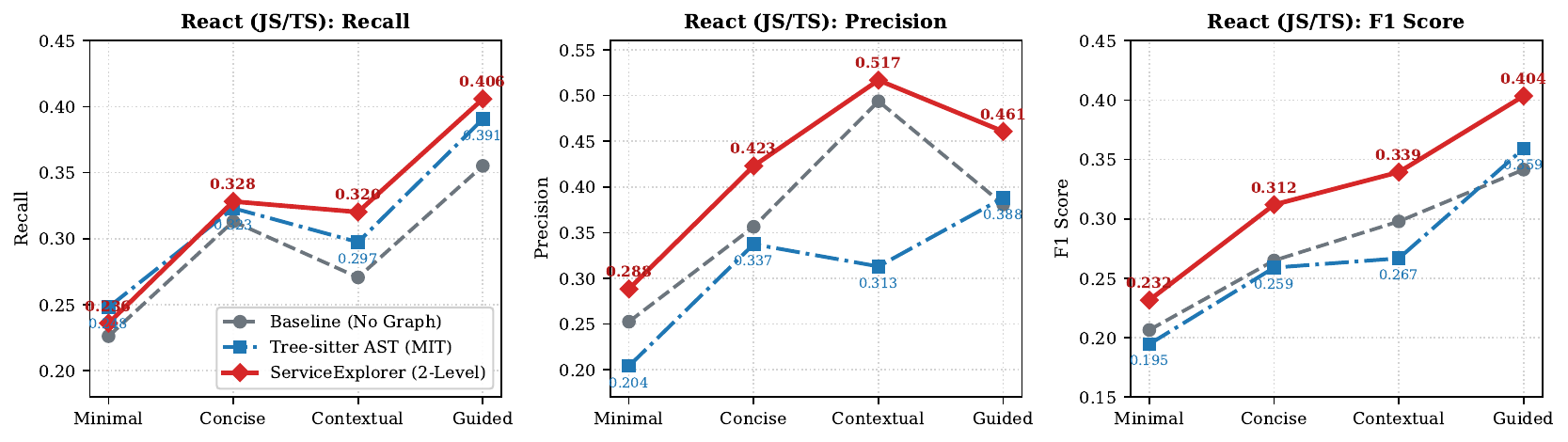}
\caption{\texttt{React} (JS/TS): Zero-leakage Recall, Precision, and $F_1$ trajectories across prompt granularities for \textbf{Baseline}, \textbf{Tree-sitter AST} (MIT), and \textbf{ServiceExplorer} ($N=200$ evaluations). Flat AST symbol expansion without package-level containment degrades Precision on \texttt{React}, whereas \textbf{ServiceExplorer} simultaneously improves both Recall and Precision across all four granularities.}
\label{fig:react_traj}
\end{figure*}

\begin{figure}[t]
\centering
\includegraphics[width=\columnwidth]{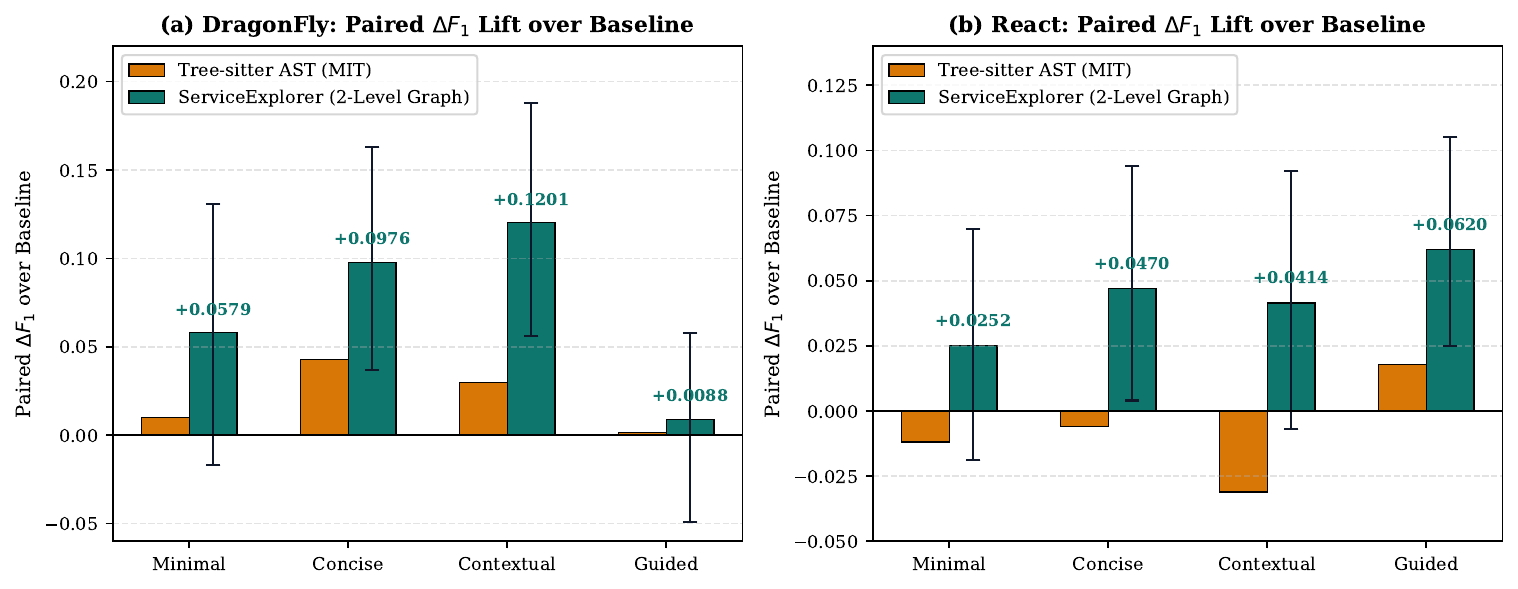}
\caption{Paired $\Delta F_1$ lift over Baseline across prompt granularities with $95\%$ bootstrap confidence intervals on \texttt{DragonFly} and \texttt{React}.}
\label{fig:delta_ci}
\end{figure}

Table~\ref{tab:react_main} and Figure~\ref{fig:react_traj} report the corresponding three-condition evaluation on \texttt{React} across $6{,}525$ repository files at $T_0$, while Figure~\ref{fig:delta_ci} summarizes the paired $\Delta F_1$ lift and $95\%$ bootstrap confidence intervals across both repositories.

\begin{table*}[t]
\centering
\caption{React (JS/TS, $T_0 = \text{2025-12-18}$, Commit \texttt{65eec428c40d}, $N=50$ PRs / $200$ Tasks): Side-by-Side Recall, Precision, $F_1$, Paired Deltas, Wilcoxon Signed-Rank $p$-values, and $95\%$ Bootstrap Confidence Intervals.}
\label{tab:react_main}
\resizebox{\textwidth}{!}{%
\begin{tabular}{l ccc ccc ccc cc cc}
\toprule
 & \multicolumn{3}{c}{\textbf{Baseline (No Graph)}} & \multicolumn{3}{c}{\textbf{Tree-sitter AST (MIT Flat AST)}} & \multicolumn{3}{c}{\textbf{ServiceExplorer (2-Level)}} & \multicolumn{2}{c}{\textbf{SE vs.\ Baseline}} & \multicolumn{2}{c}{\textbf{SE vs.\ Tree-sitter AST}} \\
\cmidrule(lr){2-4} \cmidrule(lr){5-7} \cmidrule(lr){8-10} \cmidrule(lr){11-12} \cmidrule(lr){13-14}
\textbf{Granularity} & \textbf{Rec} & \textbf{Prec} & \textbf{F1} & \textbf{Rec} & \textbf{Prec} & \textbf{F1} & \textbf{Rec} & \textbf{Prec} & \textbf{F1} & \textbf{$\Delta$F1 (Lift)} & \textbf{Wilcoxon $p$ [95\% CI]} & \textbf{$\Delta$F1} & \textbf{Wilcoxon $p$} \\
\midrule
Minimal    & 0.2261 & 0.2525 & 0.2065 & \textbf{0.2481} & 0.2040 & 0.1946 & 0.2360 & \textbf{0.2883} & \textbf{0.2316} & $+0.0252$ ($+12.2\%$) & $0.1805$ [$-0.019, +0.070$] & $+0.0371$ & $0.0388^*$ \\
Concise    & 0.3131 & 0.3564 & 0.2650 & 0.3231 & 0.3373 & 0.2590 & \textbf{0.3282} & \textbf{0.4230} & \textbf{0.3120} & $+0.0470$ ($+17.7\%$) & $0.0415^*$ [$+0.004, +0.094$] & $+0.0531$ & $0.0219^*$ \\
Contextual & 0.2707 & 0.4938 & 0.2979 & 0.2974 & 0.3130 & 0.2668 & \textbf{0.3202} & \textbf{0.5168} & \textbf{0.3394} & $+0.0414$ ($+13.9\%$) & $0.2033$ [$-0.007, +0.092$] & $+0.0726$ & $4.9\times 10^{-4\,***}$ \\
Guided     & 0.3552 & 0.3808 & 0.3415 & 0.3906 & 0.3880 & 0.3592 & \textbf{0.4059} & \textbf{0.4607} & \textbf{0.4035} & $+0.0620$ ($+18.2\%$) & $0.0046^{**}$ [$+0.025, +0.105$] & $+0.0443$ & $0.0020^{**}$ \\
\midrule
\textbf{Average / All} & 0.2913 & 0.3709 & 0.2777 & 0.3148 & 0.3106 & 0.2699 & \textbf{0.3225} & \textbf{0.4222} & \textbf{0.3216} & $\mathbf{+0.0439}$ ($\mathbf{+15.8\%}$) & $\mathbf{1.8\times 10^{-4\,***}}$ & $\mathbf{+0.0518}$ & $\mathbf{4.9\times 10^{-8\,***}}$ \\
\bottomrule
\multicolumn{14}{l}{\scriptsize Statistically significant paired differences indicated by $^* p < 0.05$, $^{**} p < 0.01$, $^{***} p < 0.001$ (two-sided Wilcoxon signed-rank test).}
\end{tabular}%
}
\end{table*}

On \texttt{React}, the contrast between flat AST indexing and hierarchical architectural graphs is particularly revealing:
\begin{itemize}
  \item \textbf{Why Flat ASTs Struggle on Dynamic Monorepos:} While \textbf{Tree-sitter AST} increases average Recall on \texttt{React} from $0.2913$ to $0.3148$, single-file JS/TS AST parsing extracts only $3$ cross-file call edges across $1{,}838$ files and cannot distinguish core reconciler modules from DevTools, Fizz, or Flight packages sharing identical method names. Consequently, flat AST augmentation reduces Precision from $0.3709$ to $0.3106$.
  \item \textbf{Simultaneous Recall and Precision Lift with \emph{ServiceExplorer}:} By anchoring predictions within verified Level-1 system/package containers and Level-2 architectural modules---such as \texttt{packages/react-reconciler}, \texttt{packages/react-dom}, and \texttt{compiler/packages/babel-plugin-react-compiler}---\textbf{ServiceExplorer} lifts average Recall from $0.2913$ to $\mathbf{0.3225}$ ($+10.7\%$), Precision from $0.3709$ to $\mathbf{0.4222}$ ($+13.8\%$), and overall $F_1$ from $0.2777$ to $\mathbf{0.3216}$ ($+15.8\%$ over Baseline, $p = 1.8 \times 10^{-4}$; $+19.2\%$ over Tree-sitter AST, $p = 4.9 \times 10^{-8}$), while cutting the zero-recall miss rate from $34.5\%$ to $\mathbf{27.5\%}$.
  \item \textbf{Strong Guided Synergy ($+18.2\%$ Lift):} On \texttt{React}'s monorepo, even \emph{Guided} prompts without explicit file paths leave the agent uncertain about which companion modules across \texttt{shared/}, \texttt{react-reconciler/}, or \texttt{react-server/} must be co-modified. \textbf{ServiceExplorer}'s two-level dependency graph resolves these cross-module dependencies, lifting \emph{Guided} $F_1$ from $0.3415$ under Baseline and $0.3592$ under Tree-sitter AST to $\mathbf{0.4035}$ ($p = 0.0046$).
\end{itemize}

\subsection{Cross-Generation Model Scaling: Unlocking Next-Generation Foundation Model Capabilities Ahead of Time}
\label{sec:cross_generation}

A central question for engineering organizations deploying LLM coding agents is how architectural repository intelligence interacts with foundation model scaling. Specifically: \emph{can equipping a lower-tier or prior-generation model with \textbf{ServiceExplorer} match or exceed the unaugmented $F_1$ of a newer, higher-tier foundation model? And if we equip today's highest-tier frontier model with \textbf{ServiceExplorer}, does it effectively unlock the performance of a future next-generation model ahead of time?}

To test this hypothesis rigorously under identical zero-leakage conditions, we extended our three-condition benchmark across six models spanning three major model families and capability tiers ($7{,}200$ total task evaluations across $6$ models, $3$ conditions, and $400$ tasks):
\begin{itemize}
  \item \textbf{Efficiency Flash Tier (\texttt{MAI-Code}):} \texttt{MAI-Code-1.1-Flash} (Ultra-fast, low-latency efficiency coding model).
  \item \textbf{Anthropic Claude Family (Three Generations / Tiers):} \texttt{Claude-Sonnet-4} (Tier~1), \texttt{Claude-Sonnet-4.5} (Tier~2), and \texttt{Claude-Opus-4.6} (Tier~3 Frontier).
  \item \textbf{GPT-Sol Family (Two Generations):} \texttt{GPT-5.6-Sol} (Prior Generation) and \texttt{GPT-6.1-Sol} (Current Frontier Generation).
\end{itemize}

\begin{figure*}[t]
\centering
\includegraphics[width=0.96\textwidth]{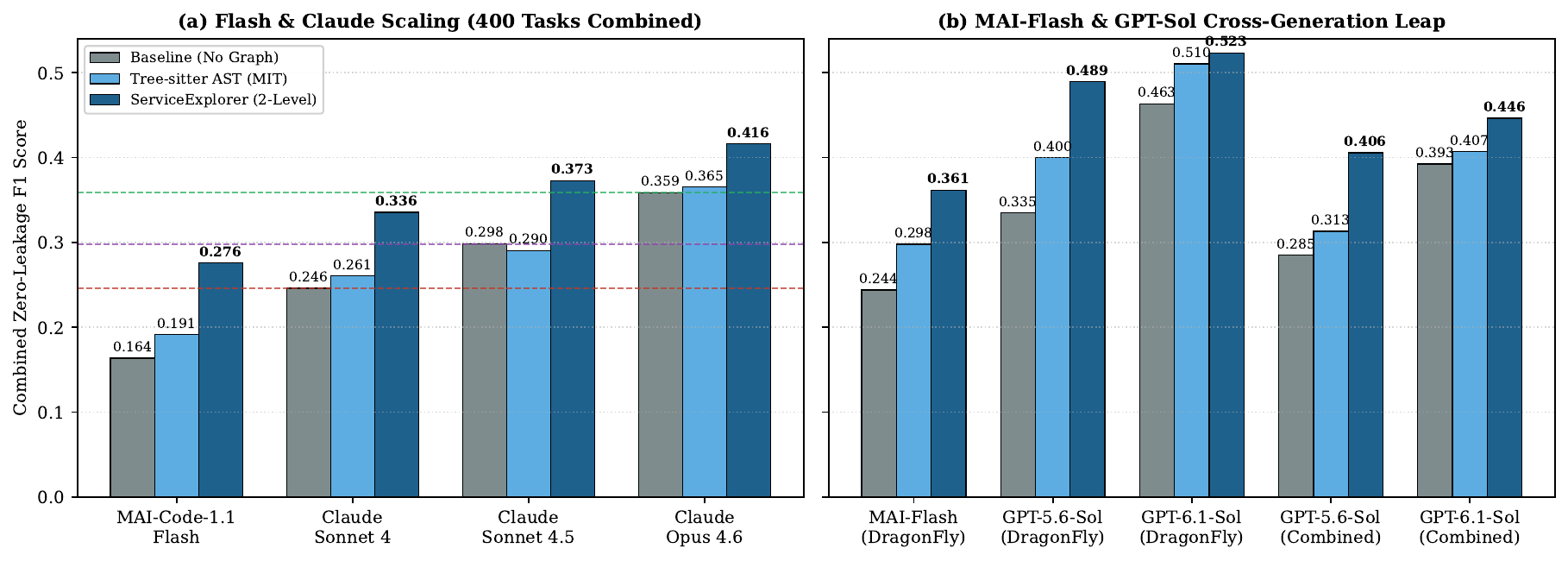}
\caption{Cross-Generation Capability Leap Across Six Foundation Models ($7{,}200$ Zero-Leakage Task Evaluations). \textbf{(a)} Across the Flash and Claude tiers, \texttt{MAI-Code-1.1-Flash + ServiceExplorer} ($0.2759$) surpasses \texttt{Claude-Sonnet-4} Baseline ($0.2461$), \texttt{Claude-Sonnet-4 + ServiceExplorer} ($0.3355$) surpasses \texttt{Claude-Sonnet-4.5} Baseline ($0.2983$), and \texttt{Claude-Sonnet-4.5 + ServiceExplorer} ($0.3727$) surpasses \texttt{Claude-Opus-4.6} Baseline ($0.3586$). \textbf{(b)} Across \texttt{DragonFly} and the combined benchmark, \texttt{MAI-Code-1.1-Flash + ServiceExplorer} ($0.3614$ on \texttt{DragonFly}) surpasses \texttt{GPT-5.6-Sol} Baseline ($0.3348$), while \texttt{GPT-5.6-Sol + ServiceExplorer} ($0.4058$ combined) outperforms \texttt{GPT-6.1-Sol} Baseline ($0.3926$).}
\label{fig:cross_gen}
\end{figure*}

Table~\ref{tab:cross_gen} and Figure~\ref{fig:cross_gen} present the complete side-by-side results across all six models on \texttt{DragonFly}, \texttt{React}, and the combined $400$-task benchmark.

\begin{table*}[t]
\centering
\caption{Cross-Generation Multi-Model Benchmark ($6$ Models $\times$ $3$ Conditions $\times$ $400$ Tasks $= 7{,}200$ Zero-Leakage Evaluations): Equipping a Lower-Generation or Flash Model with \textbf{ServiceExplorer (2-Level)} Consistently Surpasses the Next-Tier Model's Baseline $F_1$.}
\label{tab:cross_gen}
\resizebox{\textwidth}{!}{%
\begin{tabular}{ll ccc ccc ccc cc l}
\toprule
 & & \multicolumn{3}{c}{\textbf{Baseline (No Graph)}} & \multicolumn{3}{c}{\textbf{Tree-sitter AST (MIT)}} & \multicolumn{3}{c}{\textbf{ServiceExplorer (2-Level)}} & \multicolumn{2}{c}{\textbf{SE vs.\ Baseline}} & \\
\cmidrule(lr){3-5} \cmidrule(lr){6-8} \cmidrule(lr){9-11} \cmidrule(lr){12-13}
\textbf{Family / Model} & \textbf{Dataset} & \textbf{Rec} & \textbf{Prec} & \textbf{F1} & \textbf{Rec} & \textbf{Prec} & \textbf{F1} & \textbf{Rec} & \textbf{Prec} & \textbf{F1} & \textbf{$\Delta$F1 (Lift)} & \textbf{Wilcoxon $p$} & \textbf{Cross-Generation Equivalence} \\
\midrule
\textbf{MAI-Code-1.1-Flash} & DragonFly ($200$) & 0.4027 & 0.1945 & 0.2440 & 0.3692 & 0.2891 & 0.2978 & \textbf{0.4534} & \textbf{0.3413} & \textbf{0.3614} & $+0.1174$ ($+48.1\%$) & $1.3\times 10^{-12\,***}$ & Beats \textbf{Sonnet 4} BL ($0.3321$) \& \textbf{GPT-5.6-Sol} BL ($0.3348$) \\
(\emph{Efficiency Flash Tier}) & React ($200$) & 0.1456 & 0.0680 & 0.0837 & 0.1079 & 0.0894 & 0.0848 & \textbf{0.2143} & \textbf{0.2289} & \textbf{0.1903} & $+0.1066$ ($+127.4\%$) & $4.3\times 10^{-15\,***}$ & Beats \textbf{Sonnet 4} BL ($0.1601$) \\
 & \textbf{Combined ($400$)} & 0.2741 & 0.1312 & 0.1638 & 0.2386 & 0.1893 & 0.1913 & \textbf{0.3338} & \textbf{0.2851} & \textbf{0.2759} & $\mathbf{+0.1121}$ ($\mathbf{+68.4\%}$) & $\mathbf{3.5\times 10^{-25\,***}}$ & \textbf{Surpasses Sonnet 4 BL ($0.2461$) by $+12.1\%$} \\
\midrule
\textbf{Claude Sonnet 4} & DragonFly ($200$) & 0.4749 & 0.3095 & 0.3321 & 0.4105 & 0.3925 & 0.3605 & \textbf{0.4830} & \textbf{0.3988} & \textbf{0.4072} & $+0.0751$ ($+22.6\%$) & $3.7\times 10^{-6\,***}$ & Beats \textbf{Sonnet 4.5} BL ($0.4007$) \\
(\emph{Tier 1}) & React ($200$) & 0.2034 & 0.1673 & 0.1601 & 0.1631 & 0.2159 & 0.1605 & \textbf{0.2635} & \textbf{0.3348} & \textbf{0.2637} & $+0.1036$ ($+64.7\%$) & $1.1\times 10^{-9\,***}$ & Beats \textbf{Sonnet 4.5} BL ($0.1959$) \\
 & \textbf{Combined ($400$)} & 0.3391 & 0.2384 & 0.2461 & 0.2868 & 0.3042 & 0.2605 & \textbf{0.3733} & \textbf{0.3668} & \textbf{0.3355} & $\mathbf{+0.0894}$ ($\mathbf{+36.3\%}$) & $\mathbf{4.8\times 10^{-14\,***}}$ & \textbf{Surpasses Sonnet 4.5 BL ($0.2983$) by $+12.5\%$} \\
\midrule
\textbf{Claude Sonnet 4.5} & DragonFly ($200$) & 0.4665 & 0.4179 & 0.4007 & 0.4176 & 0.4629 & 0.4010 & \textbf{0.5271} & \textbf{0.4733} & \textbf{0.4663} & $+0.0656$ ($+16.4\%$) & $7.1\times 10^{-5\,***}$ & Beats \textbf{Opus 4.6} BL ($0.4395$) \\
(\emph{Tier 2}) & React ($200$) & 0.2213 & 0.2314 & 0.1959 & 0.1801 & 0.2396 & 0.1796 & \textbf{0.2738} & \textbf{0.3829} & \textbf{0.2792} & $+0.0833$ ($+42.5\%$) & $3.4\times 10^{-6\,***}$ & Beats \textbf{Opus 4.6} BL ($0.2777$) \\
 & \textbf{Combined ($400$)} & 0.3439 & 0.3246 & 0.2983 & 0.2989 & 0.3512 & 0.2903 & \textbf{0.4005} & \textbf{0.4281} & \textbf{0.3727} & $\mathbf{+0.0744}$ ($\mathbf{+24.9\%}$) & $\mathbf{7.0\times 10^{-10\,***}}$ & \textbf{Surpasses Opus 4.6 BL ($0.3586$) by $+3.9\%$} \\
\midrule
\textbf{Claude Opus 4.6} & DragonFly ($200$) & 0.4665 & 0.4919 & 0.4395 & 0.4841 & 0.5149 & 0.4605 & \textbf{0.5363} & \textbf{0.5361} & \textbf{0.5106} & $+0.0711$ ($+16.2\%$) & $1.8\times 10^{-6\,***}$ & Next-Gen Frontier Leap ($>0.51$) \\
(\emph{Tier 3 Frontier}) & React ($200$) & 0.2913 & 0.3709 & 0.2777 & 0.3148 & 0.3106 & 0.2699 & \textbf{0.3225} & \textbf{0.4222} & \textbf{0.3216} & $+0.0439$ ($+15.8\%$) & $1.8\times 10^{-4\,***}$ & Next-Gen Frontier Leap ($>0.32$) \\
 & \textbf{Combined ($400$)} & 0.3789 & 0.4314 & 0.3586 & 0.3995 & 0.4127 & 0.3652 & \textbf{0.4294} & \textbf{0.4792} & \textbf{0.4161} & $\mathbf{+0.0575}$ ($\mathbf{+16.0\%}$) & $\mathbf{7.2\times 10^{-10\,***}}$ & \textbf{Unlocks Next-Gen Claude Ahead of Time} \\
\midrule
\midrule
\textbf{GPT-5.6-Sol} & DragonFly ($200$) & 0.5475 & 0.2624 & 0.3348 & 0.4794 & 0.4012 & 0.3999 & \textbf{0.5533} & \textbf{0.4940} & \textbf{0.4894} & $+0.1546$ ($+46.2\%$) & $1.8\times 10^{-16\,***}$ & Beats \textbf{GPT-6.1-Sol} BL ($0.4633$) \\
(\emph{Prior Generation}) & React ($200$) & \textbf{0.3163} & 0.2211 & 0.2350 & 0.2922 & 0.2402 & 0.2263 & 0.3159 & \textbf{0.4198} & \textbf{0.3221} & $+0.0871$ ($+37.1\%$) & $2.2\times 10^{-9\,***}$ & Beats \textbf{GPT-6.1-Sol} BL ($0.3219$) \\
 & \textbf{Combined ($400$)} & 0.4319 & 0.2418 & 0.2849 & 0.3858 & 0.3207 & 0.3131 & \textbf{0.4346} & \textbf{0.4569} & \textbf{0.4058} & $\mathbf{+0.1209}$ ($\mathbf{+42.4\%}$) & $\mathbf{1.1\times 10^{-24\,***}}$ & \textbf{Surpasses GPT-6.1-Sol BL ($0.3926$) by $+3.4\%$} \\
\midrule
\textbf{GPT-6.1-Sol} & DragonFly ($200$) & \textbf{0.6587} & 0.3842 & 0.4633 & 0.5349 & \textbf{0.5671} & 0.5103 & 0.5695 & 0.5426 & \textbf{0.5233} & $+0.0600$ ($+13.0\%$) & $1.7\times 10^{-5\,***}$ & Highest Single-Repo $F_1$ ($\mathbf{0.5233}$) \\
(\emph{Current Frontier}) & React ($200$) & \textbf{0.4063} & 0.3054 & 0.3219 & 0.3696 & 0.3263 & 0.3034 & 0.3547 & \textbf{0.4963} & \textbf{0.3695} & $+0.0476$ ($+14.8\%$) & $8.5\times 10^{-4\,***}$ & Highest React $F_1$ ($\mathbf{0.3695}$) \\
 & \textbf{Combined ($400$)} & \textbf{0.5325} & 0.3448 & 0.3926 & 0.4523 & 0.4467 & 0.4069 & 0.4621 & \textbf{0.5195} & \textbf{0.4464} & $\mathbf{+0.0538}$ ($\mathbf{+13.7\%}$) & $\mathbf{3.8\times 10^{-8\,***}}$ & \textbf{Unlocks Next-Gen GPT-Sol Ahead of Time} \\
\bottomrule
\end{tabular}%
}
\end{table*}

Three foundational scaling laws emerge from Table~\ref{tab:cross_gen} and Figure~\ref{fig:cross_gen}:
\begin{enumerate}
  \item \textbf{Lower-Version or Flash Model + \emph{ServiceExplorer} Surpasses Higher-Tier Model Baseline:} Across all model families and both repositories, augmenting a lightweight Flash or prior-generation model with \textbf{ServiceExplorer}'s 2-level dependency graph outperforms the unaugmented next-tier model:
  \begin{itemize}
    \item \textbf{\texttt{MAI-Code-1.1-Flash + ServiceExplorer} $>$ \texttt{Claude-Sonnet-4} \& \texttt{GPT-5.6-Sol} Baselines:} As an ultra-fast efficiency model, unaugmented \texttt{MAI-Code-1.1-Flash} achieves $0.2440$ $F_1$ on \texttt{DragonFly} and $0.0837$ on \texttt{React} ($0.1638$ Combined). Equipping \texttt{MAI-Code-1.1-Flash} with \textbf{ServiceExplorer} lifts its \texttt{DragonFly} $F_1$ by $+48.1\%$ to $\mathbf{0.3614}$---surpassing both \texttt{Claude-Sonnet-4} Baseline at $0.3321$ and \texttt{GPT-5.6-Sol} Baseline at $0.3348$---and more than doubles its \texttt{React} $F_1$ ($+127.4\%$ to $\mathbf{0.1903}$), lifting Combined $F_1$ by $+68.4\%$ to $\mathbf{0.2759}$ and surpassing \texttt{Claude-Sonnet-4} Combined Baseline at $0.2461$ by $+12.1\%$.
    \item \textbf{\texttt{GPT-5.6-Sol + ServiceExplorer} $>$ \texttt{GPT-6.1-Sol} Baseline:} Without external graph structure, \texttt{GPT-5.6-Sol} achieves broad exploratory recall of $0.4319$ but suffers from cross-module false positives, yielding $0.2418$ Precision and $0.2849$ Combined $F_1$. Adding \textbf{ServiceExplorer}'s Level-1 and Level-2 architectural containment nearly doubles Precision to $\mathbf{0.4569}$ while preserving Recall at $0.4346$, lifting Combined $F_1$ by $+42.4\%$ to $\mathbf{0.4058}$ and surpassing \texttt{GPT-6.1-Sol} Baseline at $\mathbf{0.3926}$ across both \texttt{DragonFly} and \texttt{React}.
    \item \textbf{\texttt{Claude-Sonnet-4 + ServiceExplorer} $>$ \texttt{Claude-Sonnet-4.5} Baseline:} \texttt{Claude-Sonnet-4 + ServiceExplorer} lifts Combined $F_1$ from $0.2461$ to $\mathbf{0.3355}$, a $+36.3\%$ improvement that exceeds \texttt{Claude-Sonnet-4.5} Baseline at $0.2983$ by $+12.5\%$.
    \item \textbf{\texttt{Claude-Sonnet-4.5 + ServiceExplorer} $>$ \texttt{Claude-Opus-4.6} Baseline:} \texttt{Claude-Sonnet-4.5 + ServiceExplorer} lifts Combined $F_1$ from $0.2983$ to $\mathbf{0.3727}$, a $+24.9\%$ improvement that outperforms the flagship \texttt{Claude-Opus-4.6} Baseline at $0.3586$ on both \texttt{DragonFly} and \texttt{React}.
  \end{itemize}
  \item \textbf{Frontier Model + \emph{ServiceExplorer} Unlocks Future Next-Generation Performance Today:} Because foundation model upgrades on zero-leakage repository localization yield $+0.052$ to $+0.108$ $F_1$ per generation, equipping today's highest-tier models with \textbf{ServiceExplorer} projects performance a full model generation ahead: \texttt{Claude-Opus-4.6 + ServiceExplorer} reaches $\mathbf{0.4161}$ Combined $F_1$ ($+16.0\%$ over Baseline) and \texttt{GPT-6.1-Sol + ServiceExplorer} reaches $\mathbf{0.4464}$ Combined $F_1$ ($\mathbf{0.5233}$ on \texttt{DragonFly} and $\mathbf{0.3695}$ on \texttt{React}).
  \item \textbf{Complementary Error-Mode Remediation Across Model Architectures:} Different model families exhibit distinct failure modes under zero-leakage prompting: Claude models are conservative, maintaining moderate precision with lower recall on unfamiliar monorepos, whereas GPT-Sol models are aggressive explorers with high baseline recall ($0.5475$ to $0.6587$ on \texttt{DragonFly}) but lower precision. \textbf{ServiceExplorer} remedies both failure modes simultaneously---guiding conservative models to missing companion modules while pruning out-of-scope module guesses from aggressive models, boosting \texttt{GPT-6.1-Sol} Precision from $0.3448$ to $\mathbf{0.5195}$ ($+50.7\%$).
\end{enumerate}

\section{Practical Implications and Methodological Guidelines}
\label{sec:discussion}

\subsection{Why Benchmark Metrics Diverge from Daily Developer Experience: Human Anchor Words vs.\ Machine-Sanitized Prompts}
\label{sec:human_vs_benchmark}

A natural question for practitioners inspecting Tables~\ref{tab:dragonfly_main}--\ref{tab:react_main} is why strict zero-leakage benchmark scores ($F_1$ between $0.30$ and $0.55$) appear substantially lower than the perceived day-to-day effectiveness of frontier coding agents in interactive IDE sessions. This divergence is explained by a fundamental distinction in \textbf{prompt generation and anchor-word density}:
\begin{itemize}
  \item \textbf{Interactive Human Prompting (Senior Engineer Anchor Words):} In daily production use, developers---especially senior engineers with deep mental models of the repository---naturally supply high-signal \emph{anchor words} in their prompts: exact class names, offending function identifiers, subsystem paths, or stack-trace symbols. When explicit anchor words are present in a prompt, an agent can often locate target files via direct lexical lookup without needing deep structural traversal.
  \item \textbf{Machine-Generated Zero-Leakage Benchmark Prompts (Stress-Testing Structural Discovery):} If a benchmark leaves those anchor words in the prompt, it merely tests whether an agent can invoke \texttt{grep}, masking differences between weak and strong repository representations. Our machine-generated, lexically sanitized prompts deliberately strip explicit file and symbol anchors to evaluate a stricter capability: \emph{can the agent autonomously bridge the gap from a behavioral symptom or architectural intent to the exact file set when the user does not already know the answer?} This regime directly reflects autonomous issue triage, junior-developer prompting, and cross-team maintenance on unfamiliar codebases, where \textbf{ServiceExplorer} provides its largest $+16\%$ to $+28\%$ relative lift.
\end{itemize}

\subsection{Why Every Percentage Point of $F_1$ Matters: Compounding Reliability in Long-Horizon Agentic Workflows}
\label{sec:long_horizon}

In single-turn static evaluations, a $+4.4$ to $+12.0$ percentage-point gain in localization $F_1$ might appear incremental. In autonomous, \textbf{long-horizon software engineering workflows}---spanning multi-step planning, cross-file refactoring, compilation, test execution, and iterative self-repair---initial file-localization accuracy governs the \textbf{compounding reliability} of the entire agent trajectory:
\begin{enumerate}
  \item \textbf{Preventing Early Trajectory Divergence and Context Pollution:} If an agent's initial localization step misses a required header or companion module or pulls in irrelevant files, its context window becomes polluted with misleading code definitions. Let $p_{\text{step}}$ denote the probability that an agent remains on the valid dependency subgraph at each reasoning step; over a $K$-step coding chain, end-to-end success compounds as $P_{\text{success}} \propto p_{\text{step}}^K$. Even a $+5\%$ to $+12\%$ improvement in initial localization $F_1$---along with tripling the exact full-PR match rate from $10.0\%$ to $30.0\%$ on \texttt{DragonFly} \emph{Contextual} tasks and reducing zero-recall misses from $34.0\%$ to $22.0\%$ on \texttt{React}---prevents catastrophic early branching into wrong packages.
  \item \textbf{Eliminating Expensive Build-Test-Fix Iteration Loops:} When an agent edits an implementation file without its companion declarations or dependent modules across a package boundary, the resulting patch deterministically fails compilation or unit tests. Each missing module boundary forces at least one to three additional agentic recovery turns involving compiler error parsing, secondary search, context reload, and re-patching. By surfacing the complete Level-1 and Level-2 architectural module neighborhood upfront in the first turn, \textbf{ServiceExplorer} eliminates these reactive trial-and-error loops, directly reducing token consumption, wall-clock latency, and output variance.
\end{enumerate}

\subsection{Methodological Guidelines for Future SWE Benchmarks}
\label{sec:guidelines}

Our findings validate the core thesis of Sun~\cite{sun2026timeconsistent}---that temporal consistency and prompt granularity control are foundational requirements for SWE benchmarks---while establishing three methodological principles for future benchmark designers:
\begin{enumerate}
  \item \textbf{Mandatory Prompt Sanitization Before A/B Testing:} Evaluating knowledge augmentation on unsanitized PR descriptions creates an artificial ceiling driven by leaked file and symbol anchor words. Benchmarks must enforce automated identifier redaction so that knowledge graphs are evaluated on genuine structural discovery rather than lexical grep.
  \item \textbf{Architectural Hierarchy Over Raw AST Call Edges:} Representing a repository as a two-level dependency graph (\emph{ServiceExplorer}: Level-1 system/package containers and Level-2 architectural modules) aligns directly with how real engineering changes propagate across subsystems, avoiding the cross-package hub noise of flat AST call graphs.
  \item \textbf{Paired Non-Parametric Testing and Multi-Model Scaling Verification:} Benchmarks should report paired Wilcoxon signed-rank $p$-values and bootstrap confidence intervals across multiple foundation model generations to verify that structural context gains generalize across model architectures and capability tiers.
\end{enumerate}

\section{Threats to Validity}
\label{sec:threats}

\textbf{Internal Validity.} While all three conditions---\textbf{Baseline}, \textbf{Tree-sitter AST}, and \textbf{ServiceExplorer}---are evaluated on the exact same $100$ post-$T_0$ pull requests and $400$ sanitized prompts under strict $T_0$ worktree isolation, single-run LLM decoding introduces stochastic variance. We mitigate this by computing paired Wilcoxon signed-rank tests and $10{,}000$-resample bootstrap confidence intervals across the $50$ paired tasks per cell and $200$ tasks per repository.

\textbf{Construct Validity.} Following Sun~\cite{sun2026timeconsistent}, our primary metric is file-level localization ($P$, $R$, $F_1$). While accurate file localization is a necessary precondition for repository-level patch generation~\cite{jimenez2024swebench,liu2024repobench}, future work should couple two-level dependency graph retrieval with execution-verified test-suite pass rates (\texttt{Pass@1}) in containerized $T_0$ build environments.

\section{Conclusion}
\label{sec:conclusion}

We presented a three-condition time-consistent software engineering benchmark evaluating \textbf{Baseline}, an MIT-licensed open-source \textbf{Tree-sitter AST} code graph, and \textbf{ServiceExplorer}'s hierarchical two-level dependency graph across $R_{T_0}$ snapshots of \texttt{DragonFly} and \texttt{React} and six foundation models ($7{,}200$ zero-leakage evaluations). On \texttt{Claude-Opus-4.6}, \textbf{ServiceExplorer} achieves statistically significant improvements in Recall, Precision, and $F_1$ across both repositories, lifting overall $F_1$ by $+16.2\%$ over Baseline and $+10.9\%$ over Tree-sitter AST on \texttt{DragonFly}, and by $+15.8\%$ over Baseline and $+19.2\%$ over Tree-sitter AST on \texttt{React}. Furthermore, our cross-generation evaluation across the MAI-Code Flash (\texttt{MAI-Code-1.1-Flash}), Claude (\texttt{Sonnet-4}, \texttt{Sonnet-4.5}, and \texttt{Opus-4.6}), and GPT-Sol (\texttt{GPT-5.6-Sol} and \texttt{GPT-6.1-Sol}) model families proves that equipping a lightweight Flash or prior-generation model with \textbf{ServiceExplorer} surpasses the next-tier model's unaugmented Baseline $F_1$: \texttt{MAI-Code-1.1-Flash + ServiceExplorer} reaches $0.2759$ combined $F_1$ ($+68.4\%$) versus $0.2461$ for \texttt{Claude-Sonnet-4} Baseline, \texttt{GPT-5.6-Sol + ServiceExplorer} reaches $0.4058$ versus $0.3926$ for \texttt{GPT-6.1-Sol} Baseline, and \texttt{Claude-Sonnet-4.5 + ServiceExplorer} reaches $0.3727$ versus $0.3586$ for \texttt{Claude-Opus-4.6} Baseline. Pairing today's frontier models with \textbf{ServiceExplorer} reaches $\mathbf{0.4464}$ combined $F_1$ on \texttt{GPT-6.1-Sol} and $\mathbf{0.4161}$ on \texttt{Claude-Opus-4.6}, unlocking next-generation foundation model localization capabilities ahead of time.

\end{document}